\documentclass[aip,pop,reprint]{revtex4-1}
\usepackage{graphics}
\usepackage{graphicx}
\usepackage{epstopdf}
\usepackage{hyperref}

\draft 

\begin{document}


\title{Liquid crystal films as on-demand, variable thickness (50-5000 nm) targets for intense lasers} 



\author{P. L. Poole}
\email[]{poole.134@osu.edu}
\homepage[]{http://hedp.osu.edu/}
\author{C. D. Andereck}
\author{D. W. Schumacher}
\author{R. L. Daskalova}
\author{S. Feister}
\author{K. M. George}
\author{C. Willis}
\author{K. U. Akli}
\author{E. A. Chowdhury}
\affiliation{Physics Department, The Ohio State University, Columbus, Ohio 43210, USA}


\date{\today}

\begin{abstract}
We have developed a new type of target for intense laser-matter experiments that offers significant advantages over those currently in use. The targets consist of a liquid crystal film freely suspended within a metal frame. They can be formed rapidly on-demand with thicknesses ranging from nanometers to micrometers, where the particular value is determined by the liquid crystal temperature and initial volume as well as by the frame geometry. The liquid crystal used for this work, 8CB (4'-octyl-4-cyanobiphenyl), has a vapor pressure below $10^{-6}$ Torr, so films made at atmospheric pressure maintain their initial thickness after pumping to high vacuum. Additionally, the volume per film is such that each target costs significantly less than one cent to produce. The mechanism of film formation and relevant physics of liquid crystals are described, as well as ion acceleration data from the first shots on liquid crystal film targets at the Ohio State University Scarlet laser facility.
\end{abstract}


\maketitle 

\section{Introduction}

The past decade has seen the rapid development of new lasers with peak intensities in excess of  $10^{20}$ W/cm$^2$. These intensities enable a wide range of applications, including ion acceleration and its use in neutron radiography, \cite{Roth13} ion cancer therapy, \cite{Bulanov04} and nuclear physics, \cite{Santala01} as well as energetic electron acceleration to generate both x-rays \cite{Murnane89, Kneip09} for advanced imaging and positron beams \cite{Chen09} for laboratory astrophysics.

As high-intensity laser technology has developed, so too have the targets used in these experiments. It is now recognized that targets can be optimized for differing acceleration mechanisms by varying their thickness. Target Normal Sheath Acceleration (TNSA) \cite{Hatchett00, Snavely00} is the dominant ion acceleration process for few-$\mu$m-thick targets, while decreasing the thickness to sub-$\mu$m levels will unlock Radiation Pressure Acceleration (RPA) \cite{Esirkepov03} and in turn the Break-Out Afterburner (BOA) \cite{Yin06} regimes, each of which has characteristic ion signals that are useful for different applications. The target thickness range where these acceleration mechanisms turn on or dominate can vary depending on laser energy and intensity, and a target with finely-tunable thickness will allow acceleration optimization studies with unprecedented ease.

Liquid targets have been investigated previously for High Energy Density Physics (HEDP) applications. Liquid droplets or jets have been used to facilitate high repetition rate studies, \cite{Karsch03, Ter-Avetisyan06} where they were typically ejected from piezo-controlled nozzles in the target chamber. Though some effort has been devoted to making sub-$\mu$m droplet sprays with high number density, \cite{Ter-Avetisyan03} the dispersed nature of the droplet cloud decreases the effective interaction region with the laser. Additionally, the three-dimensional expansion of the droplet can lead to isotropic acceleration and faster plasma cooling, resulting in lower ion energies than planar targets. \cite{Schnurer07, Henig09} Furthermore, the relatively high vapor pressure of water (around 20 Torr) requires at least a differential pumping scheme to protect critical optical components within the target chamber and significantly restricts high-intensity short pulse experimentation due to phase nonlinearities that will accumulate as the beam propagates through the poor vacuum surrounding the target.

Here we present the development of and first experiments on target films consisting of a liquid crystal. These films are freely suspended within a metal frame with no substrate on either side, and can be rapidly made from bulk liquid crystal to have any desired thickness from 50 nm to over 5000 nm. With this real-time thickness variability, high vacuum compatibility, and low cost, liquid crystal films are ideal for studying single-shot, low Z ion acceleration in a planar target morphology. Section II details the design of the film target including liquid crystal chemistry and properties and the film formation apparatus and technique. Section III includes experimental results from the first shots taken with this film target at the Ohio State University Scarlet laser facility. Section IV is a summary of these results and the expected impact of this apparatus on HEDP and other intense laser experiments, including the prospect of high repetition rate liquid crystal film targets.

\section{Film target design}

\subsection{Liquid crystal chemistry}

Combining the physical properties of both liquids and solids, liquid crystals are long-chain molecules consisting chiefly of hydrocarbons. The crystal-like properties arise from phase-dependent orientational and positional order of these molecules. In general liquid crystals undergo a phase transition as they shift between different ordered states; as a result liquid crystals are often characterized by what types of order they can take on and under what conditions they undergo phase transitions between those states. It is by taking advantage of such phase transitions that we can obtain variable thickness thin films using liquid crystals.

Two common liquid crystal phases are the nematic phase (from the Greek \textit{nemos} for threadlike) where the molecules have strong orientational order but no positional order, and smectic (from the Latin \textit{smecticus} for soap-like) where the molecules exhibit both orientational and positional order. This smectic phase is optimum for making thin films, as it is comprised of stacks of thin directionally-oriented layers. The thickness of a smectic liquid crystal film is quantized by these layer sizes--about 3 nm for 8CB. \cite{IvanovBook} Additionally, the strong bonding between molecular layers results in vapor pressures well below $10^{-6}$ Torr.

4'-octyl-4-cyanobiphenyl, or 8CB, is a thermotropic liquid crystal, meaning its phase transitions occur due to temperature variations. 8CB transitions from its solid to the smectic phase around 294.5 K, from smectic to nematic at 306.5 K, and from nematic to isotropic (or conventional liquid) phase at 313.5 K.\cite{Clark93} It is convenient for thin film formation because its smectic phase lies within normal room temperatures. 8CB has a density near that of water and a viscosity that varies with temperature but is thicker than ethylene glycol in its smectic phase.

Smectic liquid crystal films with no substrate have been studied both in air and vacuum environments. \cite{Rosenblatt79, Morris90, Link96, Daya97, Pettey98, Bohley08} These freely suspended liquid crystal films (FSLC) are often of the smectic variety for their thin-film creation properties, and are studied as the carriers of various types of electrically-induced flow.\cite{Morris90} Though liquid crystals have traditionally been investigated for their behavior under the influence of electric or magnetic fields, our current interest in liquid crystals lies in their ability to form sub-micron-thick films in vacuum.

\begin{figure} [h]
	\centering
	\includegraphics[width=0.47\textwidth]{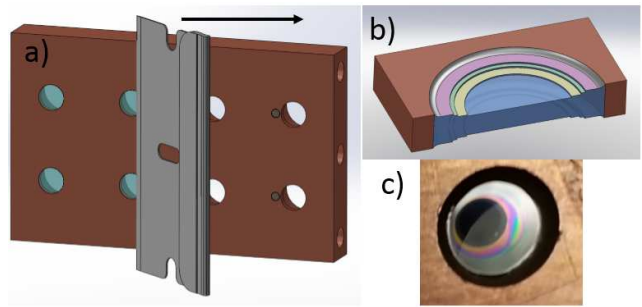}
	\caption{a) Liquid crystal film formation apparatus consisting of a copper frame with 4 mm diameter apertures. A small volume of liquid crystal placed on the edge of a wiper can be drawn across these frame apertures in a controlled fashion, leaving a liquid crystal film in its wake. b) Film morphology cartoon depicting an inner thin region surrounded by a thicker meniscus that attaches film to frame. The transition between these two thicknesses is not continuous, but rather occurs in a step-like fashion. c) Actual film showing inner transparent region surrounded by colored fringes which indicate greater thickness.}
	\label{fig:apparatus}
\end{figure}

\subsection{Film formation}

A schematic of the film generation apparatus is shown in Fig. \ref{fig:apparatus}a. Freely suspended films can be formed by drawing a precise volume of liquid crystal across an aperture in a rigid frame. A razor blade is often used for drawing, although any material that does not absorb the liquid crystal will suffice. This formation method results in a film shape as depicted in Fig. \ref{fig:apparatus}b with a thin inner region that widens near the frame edge into a thick meniscus. The transition between this thin region and the meniscus is not continuous, but rather consists of a series of steps corresponding to some multiple of the smectic layer height (3 nm for 8CB). It is not uncommon for the inner region to consist of tens of layers and the meniscus to have many hundreds, depending on the amount of liquid crystal used to make the film. Figure \ref{fig:apparatus}c shows an actual film within a 4 mm diameter frame, but liquid crystal films as large as a several cm in diameter are possible.

Film thickness is affected by the wiper angle, wiper draw speed, and frame aperture size, but we find that the dominant variables are the volume of 8CB applied and the temperature of the metal frame as the film is being drawn. Volumes as small as 100 nL are sufficient to produce a film, which results in extremely low cost per target--the bulk price of the 8CB used in our experiment was \$15/mL, resulting in a cost per target well under one cent. Without temperature control, larger initial volume results only in a thicker meniscus with no increase in the thickness of the inner region, which is typically a few hundred nm thick. 

Films around 500 nm and thicker are achievable only with temperature control of the film frame or liquid crystal volume itself. For this reason we use a copper frame heated by 25 W (max) cartridge heaters operated at low current. We have found that the ideal temperature for thick film formation lies between 27.5 $^{\circ}$C and 28.5 $^{\circ}$C, a few degrees below the smectic/nematic phase transition.  A proportional-integral-derivative (PID) controller and thermocouple maintain the desired frame temperature to within 0.1 $^{\circ}$C for optimum film formation control. At room temperature a drawn film will have a transparent inner region and an opaque white meniscus that resembles bulk liquid crystal. If the frame is heated to the appropriate temperature after a film has been drawn, the meniscus region that began opaque will become transparent and move inward to cover the inner thin layer over a period of seconds. This often results a film with thickness that varies in a step-like fashion radially increasing from the center, as depicted in Fig. \ref{fig:apparatus}b and \ref{fig:apparatus}c. Films drawn with liquid crystal and frame pre-heated to 28.0 $\pm$ 0.5 $^{\circ}$C will form with no meniscus and instead have a thick inner area. Additionally, horizontal frame orientation is optimal to prevent the heated liquid crystal from draining due to gravity, which results in a thicker film at the bottom of the frame. This draining effect can be minimized if the frame is allowed to cool to room temperature in its horizontal orientation, and is entirely eliminated if the film is also formed with no meniscus. Under these conditions, films with variable thickness can be readily formed for use in high field physics experiments.

Liquid crystal films can exhibit other thickness variations than the meniscus/thin inner region morphology. It is not uncommon in freely suspended liquid crystal films for small volumes to move away from the meniscus to float on top of the thin inner layers; the formation and manipulation of these ``islands'' are the subject of study within liquid crystal fluid dynamics.\cite{Pattanaporkratana04} They are mobile and will respond to air currents and gravity. Though they usually return eventually to the meniscus region, islands can be many layers thick and as much as a few mm across while moving over the film. Islands represent a difficulty for liquid crystal films as HEDP targets in that they constitute a large, mobile thickness variation. Additionally, the meniscus of a film drawn horizontally but then turned vertically will drain downward due to gravity over the course of minutes. While this process will not destroy the film, the target will eventually be a few hundred nm thick at its top and a few microns thick at the bottom. \textit{To address both of these problems and achieve a consistent thickness the meniscus region must be reduced or eliminated.} This can be done with careful temperature control within the previously mentioned range and by using a minimal volume of liquid crystal for each film, allowing film thicknesses between 50 nm and 5000 nm or more.

\begin{figure} [h]
	\centering
	\includegraphics[width=0.47\textwidth]{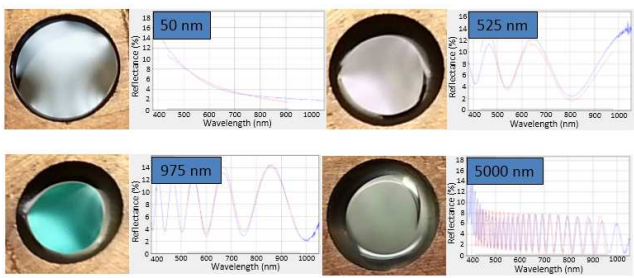}
	\caption{Example film thicknesses with measurement from the Filmetrics multi-spectral interferometer demonstrating the thickness range of these liquid crystal targets. The plots show reflectance vs. wavelength where the blue curve is the measured reflectance from the film and the red curve is the result of an iterative calculation to match this spectrum. In general, more peaks correspond to a thicker film. The images to the left of each plot show an example film within a copper frame at each of these thicknesses.}
	\label{fig:filmthick}
\end{figure}

The ability to tune thickness on-demand is a feature not present in traditional solid targets, but requires a measurement method to determine the film thickness. Because the desired smectic liquid crystals tend to be on the order of optical wavelengths and are transparent in their freely suspended smectic state, thin film interferometry can be used to measure thickness. This is the mechanism utilized in the Filmetrics F20 multi-spectral thin film measurement device, which was employed for this experiment. This device measures the reflected spectrum using a known input spectrum from 400 nm to 1000 nm (halogen lamp source) and the reflected spectrum from a reference material to determine thickness in real-time. The thickness range of 8CB is demonstrated in Fig. \ref{fig:filmthick} with films and their corresponding thickness measurements from the Filmetrics interferometer. Note that these films are of one color in contrast to the one shown in Fig. \ref{fig:apparatus}c, indicating uniform thickness.

Of note is the resiliency of liquid crystal films formed in the manner described here. A film drawn with little or no meniscus region will maintain its original thickness regardless of orientation, temperature, surrounding pressure, or gross motion of the frame. As a result the method for using these as intense laser targets is to make the film to the desired thickness using a heated frame, allow the frame to cool back to room temperature, mount the frame within the target chamber, then pump the chamber down to normal vacuum operating levels. Film thickness has been monitored carefully and observed not to fluctuate during these temperature or pressure changes. In fact, liquid crystal films brought down to 10$^{-6}$ Torr and not shot were then re-measured once the target chamber had been vented back to atmospheric pressure, at which point their thickness was found to be identical to the original value at formation.

\section{Experimental results}

\begin{figure} [h]
	\centering
	\includegraphics[width=0.47\textwidth]{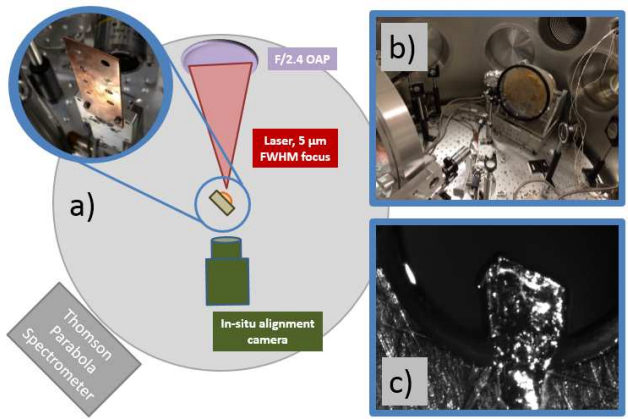}
	\caption{a) Schematic of the Scarlet laser main experimental target chamber. An F/2.4 off-axis parabola (OAP) sends the 5 inch diameter beam toward the copper target frame shown in the left inset. An in-vacuum camera situated behind the target along the laser axis is used for alignment, while a Thomson parabola spectrometer collects ion data along the rear-target-normal direction. b) The actual experimental setup with OAP in the background and the target frame and alignment objective in the foreground. c) Image of a 3 $\mu$m Al film alignment fiducial within a liquid crystal film, as seen from another in-situ camera.}
	\label{fig:chamber}
\end{figure}

To evaluate liquid crystal films for high field physics experiments we performed an initial ion acceleration experiment using these targets. The experiment was done on the Scarlet laser facility at Ohio State, which is a 400 TW Titanium:sapphire-based short-pulse laser system that delivers a maximum of 12 J of linearly-polarized 800 nm light in a 30 fs pulse to a 5 $\mu$m FWHM focal spot (F/2.4) at a repetition rate of once per minute. For these shots 4 J was delivered to the target at a 22.5$^{\circ}$ angle of incidence. The chief diagnostic was a Thomson parabola spectrometer \cite{Morrison11} located behind target normal to observe TNSA ions. A schematic of the experimental chamber setup is shown in Fig. \ref{fig:chamber}a and \ref{fig:chamber}b.

Our 8CB was obtained from Alpha Micron. An initial volume between 100 and 1000 nL was delivered to the heated target frame using a Harvard Apparatus PHD Ultra syringe pump.

The targets were aligned by floating a small 3 $\mu$m Al foil within the liquid crystal film itself, illuminating this edge with an alignment laser backlight, and imaging the shadow cast onto an in-vacuum camera setup located directly behind the target along the laser axis. An image of a film and this alignment fiducial is shown in Fig. \ref{fig:chamber}c. The camera is imaged  by an in-situ 10x infinity-corrected microscope objective with a 3.5 $\mu$m depth of focus over a 0.5 mm field of view. The camera and objective lens are situated on a vertical translation stage such that they can be moved safely away during a shot.

\begin{figure} [h]
	\centering
	\includegraphics[width=0.47\textwidth]{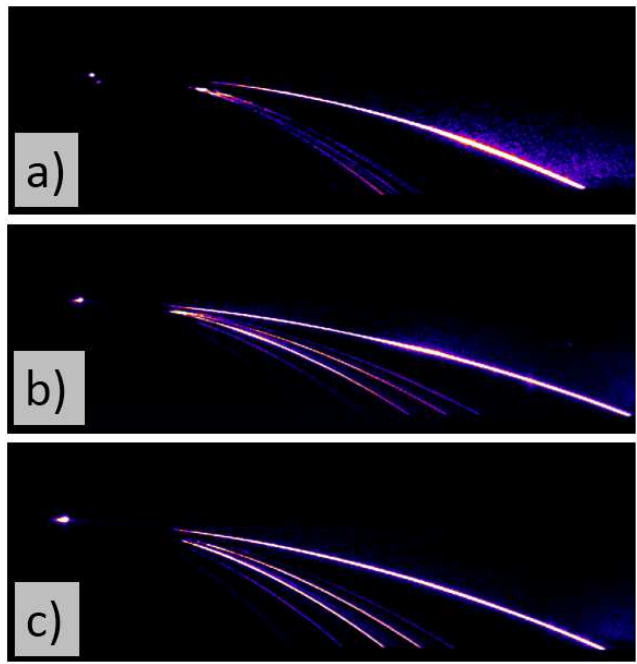}
	\caption{Ion acceleration data from the Thomson parabola spectrometer. a) shows the accelerated ions from a 100 nm Si$_3$N$_4$ solid target, b) shows the traces from a 700 nm thick liquid crystal film target, and c) is from a 160 nm film. Both liquid crystal ion traces show a strong proton signal with max energy around 10 MeV, as well as multiple traces from other ion species, and are comparable to the trace from the similar-thickness solid target.}
	\label{fig:data}
\end{figure}

Three ion traces are shown in Fig. \ref{fig:data}. The first (Fig. \ref{fig:data}a) is from a 100 nm Si$_3$N$_4$ solid target, while Fig. \ref{fig:data}b and \ref{fig:data}c are from a 700 nm thick and a 160 nm thick liquid crystal film, respectively. Maximum proton energies between 5 and 10 MeV were observed for the liquid crystal films, which is consistent with the similar-thickness Si$_3$N$_4$ targets. The other traces observed include carbon ions consistent with the known molecular makeup of 8CB.

\section{Conclusion}

We have demonstrated freely suspended liquid crystal films as a new kind of target for ultra-intense laser physics experiments. These films are stable in vacuum, inexpensive to produce, and can be varied in thickness between 50 nm and 5000 nm on-demand by simple changes in the formation process. Protons accelerated from these thin film targets at various thicknesses have been measured with a Thomson parabola spectrometer and compare favorably to similar-thickness but much more expensive solid targets. The design of this target is such that high repetition rates are possible in principle, and experiments to test this functionality are currently being designed. We believe liquid crystals can be easily and effectively applied to experiments across HEDP and other high field physics research, and they may also be useful as transparent, variable-thickness vacuum-compatible substrates for spectroscopy and similar applications.

\begin{acknowledgments}
We would like to thank Xiao-lun Wu of the University of Pittsburgh, Hiroshi Yokoyama and Peter Palffy-Muhoray of the Kent State University Liquid Crystal Institute, and Cheol Park and Kyle Meienberg from the University of Colorado for fruitful discussions, as well as Randall Hanna, Mike Prikockis, R. Sooryakumar, Jim Krygier, and R.R. Freeman of the Ohio State University for their contributions to this project. This work was supported by the DARPA PULSE program through a grant from AMRDEC and by the US Department of Energy under contract DE-NA0001976.
\end{acknowledgments}

\end{document}